\documentclass[aps,prb,twocolumn,showpacs,groupedaddress]{revtex4}
\usepackage[dvips]{graphicx}
\usepackage{amsmath}
\renewcommand{\vec}{\mathbf}
\renewcommand{\Im}{\textrm{Im}}
\begin{document}
\title{Ferromagnetism and temperature-dependent Electronic Structure of hcp Gadolinium}
\author{C.~Santos}
\affiliation{Lehrstuhl Festk{\"o}rpertheorie, Institut f{\"u}r Physik,
  Humboldt-Universit{\"a}t zu Berlin, Newtonstra{\ss}e 15, 12489 Berlin,
  Germany}
\author{V.~Eyert}
\affiliation{Institut f{\"u}r Physik, Universit{\"a}t Augsburg, Universit{\"a}tsstr.~1, 
   86135 Augsburg, Germany}
\author{W.~Nolting}
\email{Wolfgang.Nolting@physik.hu-berlin.de}
\affiliation{Lehrstuhl Festk{\"o}rpertheorie, Institut f{\"u}r Physik,
  Humboldt-Universit{\"a}t zu Berlin, Newtonstra{\ss}e 15, 12489 Berlin,
  Germany}
\date{\today}
\begin{abstract}
  We use a combination of a many-body model analysis with an ``\textit{ab initio}'' band structure
  calculation to derive the temperature dependent electronic quasiparticle structure of the
  rare-earth metal Gadolinium. As a local-moment system Gd is properly represented by the ferromagnetic
  (multiband) Kondo-lattice model (s-f (d-f) model). The single-particle part of the model-Hamiltonian 
  is taken from an augmented spherical wave (ASW) band calculation. The proposed
  method avoids the double counting of relevant interactions by exploiting an exact limiting case of 
  the model and takes into account the correct symmetry of atomic
  orbitals. The ``\textit{a priori}''
  only weakly correlated 5d conduction bands get via interband exchange coupling to the localized 4f 
  levels a distinct temperature dependence which explains by a Rudermann-Kittel-Kasuya-Yosida (RKKY) -type
  mechanism the ferromagnetism of Gd. We get a self-consistently derived Curie temperature of 294.1 K and 
  a $T=0$-moment of 7.71 $\mu_{\rm B}$, surprisingly close to the experimental values. The striking induced
  temperature-dependence of the 5d conduction bands explains respective photoemission data. 
  The only parameter of the theory (interband exchange coupling $J$) is
  uniquely fixed by the band calculation.       
\end{abstract}
\maketitle
\section{Introduction}\label{sec:intro}
The rare-earth metal Gd is one of the four elemental ferromagnetic
metals;  the others are Fe, Co, Ni.
It crystallizes in the hcp structure with a lattice constant 
$a=3.629$ \AA, $c/a=1.597$\cite{Legvold80}.
Magnetic properties result from the half-filled 4f shell ($L=0$,
$J=S=\frac{7}{2}$) which gives rise to strictly
localized magnetic moments. Conductivity properties are due to partially
filled 5d/6s conduction bands. As to
the purely magnetic properties Gd is considered an almost ideal
Heisenberg ferromagnet with a Curie temperature of
$T_{\rm C}=293.2$ K and a zero-temperature moment of
$\mu(T=0)=7.63$~$\mu_{\rm B}$ \cite{NLS63}. The  latter indicates
an induced polarization of the conduction bands of at least 0.63
$\mu_{\rm B}$ due to an interband exchange coupling
between itinerant band electrons and localized 4f electrons. The rather
strict localization of the 4f wave function\cite{HarFree74} prevents a
sufficient overlap for a direct exchange interaction between the 4f
moments. The coupling between the moments is therefore mediated by
polarized 5d/6s conduction electrons (RKKY), i.e. strongly influenced
by the electronic structure.

Although the ferromagnetic ground-state of Gd is of
course  without any doubt, it is still a matter of debate how to get
this fact by an ``\textit{ab initio}'' band structure calculation.
Numerous investigations of the electronic ground-state properties of Gd
have been performed in the recent past
\cite{Sti85,Rich89,Kru89,Tem90,Sin91,Ahu94,Sab97,Kurz02,DDN98}, all
in the framework of density functional theory (DFT). They provide a
convincing description of
ground-state properties such as the lattice constant, the hcp-crystal
stability, the $c/a$ ratio, the magnetic moment\cite{Sti85,Tem90}, the bulk
modulus, and the Fermi surface paramters\cite{Tem90,Ahu94}. On the other
hand, a standard local density approach (LDA) to DFT predicts an
antiferromagnetic ground state if the 4f electrons are considered as
valence electrons.
In a detailed analysis Kurz et al.~\cite{Kurz02} have demonstrated that
the reason for the incorrect prediction of antiferromagnetism is the
wellknown difficulty of LDA correctly to describe
strongly localized electrons. The LDA calculation of Ref.~\onlinecite{Kurz02}
poses the nearly dispersionsless majority 4f bands some 4.5 eV below the
Fermi energy while the minority 4f bands are directly above the Fermi
energy leading to a certain itinerancy of the 4f electrons. These findings
are at variance with the results of combined direct (XPS) and inverse
(BIS) photoemission experiments\cite{LBC81} which observe occupied
4f~$\uparrow$-states at the binding energy ($-7.44\pm 0.1$) eV and
unoccupied 4f~$\downarrow$-states at ($+4.04\pm0.2$) eV,
i.~e. distinctly away from the Fermi edge. A special consequence of the
wrong position of the down spin 4f states is an extremely high density
of states close to the Fermi edge and therewith an unrealistically big
$\gamma$-value of the electronic heat capacity\cite{Sti85}. The most
important consequence of the wrong 4f~$\downarrow$-position, however, is
the prediction of antiferromagnetism. All calculations, which treat the
4f electrons as valence electrons, irrespective of whether the LDA or
the generalized gradient approximation (GGA) is applied, end
up with an antiferromagnetic Gd-ground-state\cite{Kurz02,EAO95}. Today
it is clear how to remove this inadequacy of LDA (GGA) ``\textit{by
hand}''\cite{Kurz02}. One has to remove the 4f~$\downarrow$-states from
the Fermi energy. This can be done simply by considering the 4f states
as ``\textit{core states}'', so that they are not allowed to hybridize with any
other states on neighbouring atoms\cite{Kurz02,EAO95}. Another way is to
apply the so-called LDA$+U$ method\cite{AAL96}, which introduces
strong intraatomic interactions of the localized states in a
Hartree-Fock-like manner. The main effect is a splitting apart of the
occupied and unoccupied 4f states\cite{Sab97,SLP99,Kurz02}, i.e. in
particular a removal of the minority 4f states from the Fermi edge. 

\begin{figure}[ht]
\includegraphics[width=\linewidth]{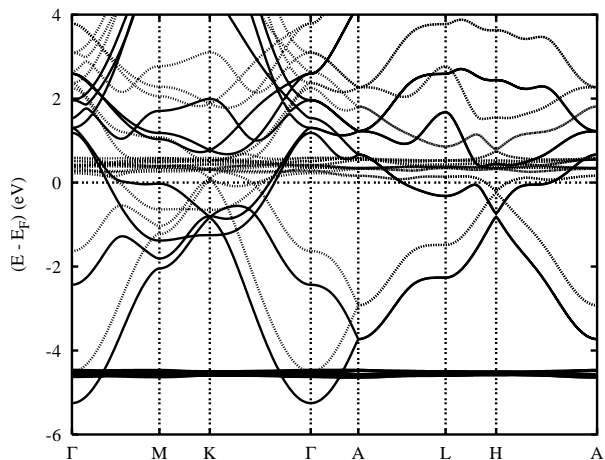}
\caption{Spin resolved $T=0$ band structure of ferromagnetic Gd as a
function of the wave vector, obtained by a scalar-relativistic ASW calculation. Solid
lines for $\uparrow$ states, dotted lines for $\downarrow$ states. The energy zero is defined 
by the Fermi edge. The flat dispersions are the 4f levels.}
\label{fig:GdBS}
\end{figure}

Needles to say that in a real ``\textit{ab initio}'' DFT-procedure all
electron states, i.e. in particular the 4f states, should be treated
as valence states. To declare the 4f states as core-states or to
introduce at a convenient place the \textit{``Hubbard-$U$''} surely
corrumpes a bit the ``\textit{first principles}'' character of the band
calculation. The only motivation is to compensate the LDA artifact
which prevents the correct ferromagnetic Gd-ground state. We have
recalculated the Gd-band structure using a new implementation of the augmented
spherical wave (ASW) method\cite{WKG79, WKG00}. The 4f electrons have
been treated as valence electrons and the scalar-relativistic
approximation of the Dirac equation has been used. Although the
antiferromagnetic configuration turns out to minimize the total energy,
the ferromagnetic order has been assumed, firstly because it is closer
to reality, secondly because we need these data for our further
procedure. The results for some high-symmetry directions are plotted in
Fig.~\ref{fig:GdBS}.
We recognize the wrong position of the rather flat 4f dispersions. The
fairly broad 5d/6s dispersions exhibit an exchange splitting with a
weak \mbox{$\vec{k}$-dependence}. The result is in good agreement with other
first-principles calculations that treat the 4f states as valence
states.\cite{Kurz02,Kru89,Sin91}. The occupied and almost dispersionless
4f~$\uparrow$-bands provide the major part of the magnetic moment
($\approx 7$ $\mu_{\rm B}$) while the 5d splitting at the Fermi edge
accounts for the experimentally observed excess moment of $\approx 0.63$
$\mu_{\rm B}$.

The mentioned 5d/6s exchange splitting must be induced since the 5d/6s
electrons can be considered as only weakly correlated and \textit{``a priori''}
non-magnetic. The splitting is obviously due to a strong interband exchange
interaction between the flat 4f states and the extended 5d/6s conduction
states. Precondition for that is a ferromagnetic order of the localized
magnetic moments built by the half-filled 4f shells. The next neighbour
distance is too large for a direct exchange interaction. The moment
coupling is of indirect nature mediated by a spinpolarization of the
conduction electrons due to the mentioned interband exchange interaction
with the localized 4f electrons. This so-called RKKY mechanism is
strongly depending on the electronic structure. To understand the
ferromagnetism of Gd does therefore mean first of all to understand its
electronic structure.

The induced 5d/6s exchange splitting is still a matter of controversial
debate, in particular what concerns its temperature
dependence\cite{DDN98}. Is it collapsing or non-collapsing for
$T\rightarrow T_{\rm C}$? Photoemission data appear to be not
unique. Some experiments point to a collapsing (``\textit{Stoner-like}'')
behavior\cite{KAE92}, others exhibit a splitting that does not shift
very much with temperature (``\textit{spin-mixing}'') persisting in the
paramagnetic phase\cite{LZD92}. In the latter case the demagnetization for
$T\rightarrow T_{\rm C}$ is reached by a redistribution of spectral
weight rather than by a gradually increasing overlap of respective spin
peaks. To find out what is really going on one needs a theory for the
full temperature-dependence of the electronic structure. Pure
``\textit{ab initio}'' band calculations are restricted to $T=0$ being
therefore insufficient for this purpose. Sandratskii and
K{\"u}bler\cite{SKU93} have proposed a DFT-based theory where
finite-temperature effects are simulated to a certain degree by a
respective directional disorder of the spatially localized 4f moments.
Even though being an interesting ansatz it certainly cannot replace the
full statistical mechanics of the local-moment ferromagnet.

A key-quantity of ferromagnetism is the Curie temperature 
$T_{\rm C}$. It is the aim of each theory for a ferromagnetic material to
approach $T_{\rm C}$ as quantitatively as possible. On the other hand,
it is a very sensitive term to get. Several attempts have been started
to estimate $T_{\rm C}$ from total-energy calculations by use of the
LDA-DFT scheme\cite{SBR02,Kurz02,TKB03}. For this purpose the energy data are
inserted into simple mean-field formula for the magnetic transition
temperature, very often arriving at astonishingly accurate $T_{\rm C}$-values.  
However, it is surely not unfair to state that such estimates cannot
replace a full theory of the Gd ferromagnetism. The latter requires
access to an electronic structure calculation which fully accounts for
decisive temperature and correlations effects. To our information such a
complete theory does not yet exist for the prototype local-moment
ferromagnet Gd. It is the aim of this paper to present a methodical
approach to the temperature-dependent electronic structure of ferromagnetic
local-moment metals with a direct application to Gd.

We present a theory of the electronic
quasiparticle structure of the ferromagnetic 4f metal Gd that yields in
a self-consistent manner the electronic as well as the magnetic
properties. The approach shall work for arbitrary temperatures regarding
in particular electron correlations effects. To get in this sense a
realistic picture of Gd we combine a ``\textit{first principles}'' band
structure calculation with a many-body evaluation of a properly chosen
theoretical model similar to previous work on the band ferromagnets Fe,
Co, and Ni\cite{NBD89}. We consider the underlying proposal a continuation
and extension of a previously pubished paper\cite{REN99} that already
dealt with the electronic quasiparticle structure of Gd. However, in the
previous case we did not succeed in getting the ferromagnetism
self-consistently, i.e. simply via the special electronic
structure. There appeared a serious ambiguity how to handle the
d-band degeneracy, i.e. how to perform the necessary decomposition of
5d-band in non-degenerate subbands. The decomposition in
Ref.~\onlinecite{REN99} did not emphasize the correct symmetry of atomic
orbitals. Separate model calculations \cite{NRM97,SNO02} revealed that the then-used d-band
decomposition is inconvenient for a ferromagnetic order of the local 4f
moments. We propose in this paper a new ansatz by which one gets
correctly the electronic structure as well as the magnetic order of Gd. 

The general procedure is briefly described in the next section. Central
part of the procedure is a many-body evaluation of a properly chosen
theoretical model. In section \ref{sec:TModel} we introduce and justify
the (multiband) Kondo-lattice model (KLM) as a good starting point for an
at least qualitative understanding of local-moment ferromagnets such as
Gd. We explain how to combine it with an LDA-DFT band calculation to
come to quantitative statements. The KLM provokes a non-trivial
many-body problem which for the general case cannot be treated
rigorously. In section \ref{sec:MBeval} our theoretical approach 
is represented. In the last step
(section~\ref{sec:MEGd}) we combine the model analysis with a band
structure calculation to get the electronic quasiparticle spectrum of Gd
and its temperature dependence, that, on the other hand, fixes the
magnetic properties of the rare earth metal as, e.g., the Curie
temperature and the magnetic moment. 
\section{General procedure}\label{sec:GProcedure}
Our study aims at a quantitative determination of the
temperature-dependent electronic structure of the 4f-ferromagnet
Gd. The general concept is rather straightforward and consists of three
steps. The important first step is the choice of a suitable theoretical
model. The main physics is due to the existence and the mutual influence
of two well defined subsystems, quasi-free electrons in rather broad
conductions bands (5d/6s) and localized electrons with extremely flat
dispersions (4f). The theoretical model is defined by its Hamiltonian;
\begin{equation}\label{eq:Thmodel}
  H=H_{0}+H_{1}
\end{equation}
more strictly, by its interaction part $H_{1}$. This particular operator
shall incorporate all those interactions which are responsible for the
characteristic phenomena under study. In the present case $H_{1}$ should
cover the decisive electron correlations which determine the magnetic
properties and the characteristic temperature dependence of the electron
quasiparticle spectrum. Our proposal for $H_{1}$ is discussed in the
next section.

While there is no contribution of the f electrons to the
kinetic energy, the part of the band electrons reads:
\begin{eqnarray}\label{eq:H0}
  H_{0}&=&\sum_{ij\sigma
    mm^{\prime}}\left(T^{mm^{\prime}}_{ij}-\mu\delta_{ij}\delta_{mm^{\prime}}\right)
c^{+}_{im\sigma}c^{}_{jm^{\prime}\sigma}\nonumber\\
&=&\sum_{\vec{k}\sigma mm^{\prime}}(T^{mm^{\prime}}_{\vec{k}}-\mu\delta_{mm^{\prime}})c^{+}_{\vec{k}m\sigma}
c^{}_{\vec{k}m^{\prime}\sigma}
\end{eqnarray}
$c^{+}_{jm\sigma}(c^{}_{jm\sigma})$ is the creation (annihilation)
operator for a Wannier electron at site $\vec{R}_{j}$ in the orbital $m$
with spin $\sigma$
($\sigma=\uparrow,\downarrow$).
$c^{+}_{\vec{k}m\sigma}(c^{}_{\vec{k}m\sigma})$
is the respective Fourier transform
\begin{equation}
c^{}_{\vec{k}m\sigma}=\frac{1}{\sqrt{N}}\sum_{j}c^{}_{jm\sigma}
\:e^{i\vec{k}\cdot\vec{R}_j}
\end{equation}
\begin{equation}
c^{}_{jm\sigma}=\frac{1}{\sqrt{N}}\sum_{\vec{k}}c^{}_{\vec{k}m\sigma}
\:e^{-i\vec{k}\cdot\vec{R}_j}
\end{equation}
The hopping process from site $\vec{R}_{j}$ to site $\vec{R}_{i}$ may be
accompanied by an orbital change ($m^{\prime}\rightarrow m$).
$T^{mm^{\prime}}_{ij}$ are the respective hopping integrals:
\begin{equation}\label{eq:HoppingFT}
  T^{mm^{\prime}}_{ij}=\frac{1}{N}\sum_{\vec{k}}T^{mm^{\prime}}_{\vec{k}}
  e^{i\vec{k}\cdot(\vec{R}_i-\vec{R}_j)}
\end{equation}
$\mu$ is the chemical potential.
The single-particle part $H_{0}$ of the model-Hamiltonian $H$ stands, as
usually, for the kinetic energy of the itinerant charge carriers and for
their interaction with the lattice potential. However, it shall
furthermore account for all those interactions which are not directly
covered by $H_{1}$. By construction these interactions are not
important for the magnetic properties and the temperature-dependence of
the electronic spectrum of the underlying ferromagnetic material
(Gd). Nevertheless they may influence the rough structure of the
spectrum so that they cannot be neglected if our study really aims at a
quantitative description of Gd. For this reason we perform in the second
step of our procedure a full selfconsistent band structure calculation within the
LDA-DFT scheme in order to replace the single-particle energies
$T^{mm^{\prime}}_{\vec{k}}$ in $H_{0}$ by the effective energies of the
LDA:
\begin{equation}\label{eq:HoppingRen}
  T^{mm^{\prime}}_{\vec{k}}\longrightarrow  
  T^{mm^{\prime}}_{\vec{k}}(\textrm{LDA})
\end{equation}

Since the \textit{``other''} interactions are, by construction of the
model, not responsible for the temperature effects, we can incorporate
them for any temperature, e.g. for $T=0$ where LDA works. It is therefore
guaranteed that all the other interactions are well accounted by the
single particle energy renormalization (\ref{eq:HoppingRen}). However,
the problem of a double counting of just the relevant interactions,
namely once explicitely in $H_{1}$ and then once more implicitely in
$H_{0}$, must carefully be avoided. How we circumvent this problem in the
case of the local-moment ferromagnet Gd is explained at a later stage.

In the third step we apply a many-body formalism in order to investigate
how the effective single-particle energies change under the influence of
the \textit{``relevant''}  interaction $H_{1}$ into temperature -, carrier
concentration ($n$) - and possibly spin - dependent quasiparticle energies:
\begin{equation}\label{eq:change}
  H_{1}: T^{mm^{\prime}}_{\vec{k}}(\textrm{LDA})\longrightarrow E_{m\sigma}(\vec{k},T,n)
\end{equation}
The quasiparticle energies are derived as the poles of the
single-electron Green-function matrix:
\begin{equation}\label{eq:GF}
  \hat G_{\vec{k}\sigma}(E)=\hbar\left[(E+\mu)\hat {\rm I}
    -\hat T_{\vec{k}}-\hat M_{\vec{k}\sigma}(E)\right]^{-1}
\end{equation}
$\hat M_{\vec{k}\sigma}(E)$ is the selfenergy matrix, the determination of
which solves the problem. $\hat T_{\vec{k}}$ is the hopping-matrix. The 
elements of the Green-function matrix are the retarded single-electron
Green functions:
\begin{equation}\label{eq:GFelements}
  \langle\langle
  c^{}_{\vec{k}m\sigma};c^{+}_{\vec{k}m^{\prime}\sigma}\rangle\rangle_{E}
=-i\int^{+\infty}_{0}dt\:e^{\frac{i}{\hbar}Et}
\langle [c^{}_{\vec{k}m\sigma}(t),c^{+}_{\vec{k}m^{\prime}\sigma}(0)]_{+}\rangle
\end{equation}
$[\dots,\dots]_{+(-)}$ means the anticommutator (commutator) and
$\langle\dots\rangle$ is the thermodynamic average. The elements of the
selfenergy matrix formally solve the Green function equation of motion:
\begin{equation}\label{eq:Fsolution}
  \langle\langle[c^{}_{\vec{k}m\sigma},H_{1}]_{-};c^{+}_{\vec{k}m^{\prime}\sigma}\rangle\rangle_{E}
\equiv\sum_{m^{\prime\prime}}M^{mm^{\prime\prime}}_{\vec{k}\sigma}(E)
\langle\langle c^{}_{\vec{k}m^{\prime\prime}\sigma};c^{+}_{\vec{k}m^{\prime}\sigma}\rangle\rangle_{E}
\end{equation}
We will discuss our results in terms of spectral densities (SD) and
quasiparticle densities of states (Q-DOS), because both have a direct
relationship to the experiment. Except for respective transition matrix
elements the spectral density expresses the bare line shape of an angle- and
spin - resolved (direct or inverse) photoemission spectrum:
\begin{eqnarray}\label{eq:Skm}
  S_{\vec{k}m\sigma}(E-\mu)&=&-\frac{1}{\pi}\Im G^{mm}_{\vec{k}\sigma}(E-\mu)\nonumber\\
  &=&-\frac{1}{\pi}\Im\langle\langle c^{}_{\vec{k}m\sigma};c^{+}_{\vec{k}m\sigma}\rangle\rangle_{E-\mu}
\end{eqnarray}
An additional $\vec{k}$-summation yields the quasiparticle density of
states (angle-averaged photoemission spectrum !)
\begin{equation}\label{eq:DOS}
  \rho_{m\sigma}(E)=\frac{1}{N\hbar}\sum_{\vec{k}}S_{\vec{k}m\sigma}(E-\mu)
\end{equation}
that in general will be temperature -, carrier concentration -, lattice
structure -, and in particular for ferromagnetic systems, explicitely
spin-dependent.
\section{Theoretical model}\label{sec:TModel}
\subsection{Model-Hamiltonian}\label{subsec:ModelH}
We still have to fix the interaction part $H_1$ of the model Hamiltonian
(\ref{eq:Thmodel}) for the correlated system of localized (4f) and
delocalized (5d) electrons. We presume from the very beginning an
on-site Coulomb interaction between electrons of different subbands,
\begin{equation}\label{eq:H1}
  H_{1}=\frac{1}{2}\sum_{L_1\cdots L_4}\sum_{\sigma\sigma^{\prime}}
  U_{L_1\cdots L_4}c^{+}_{L_1\sigma}c^{+}_{L_2\sigma^{\prime}}
  c^{}_{L_3\sigma^{\prime}}c^{}_{L_4\sigma}
\end{equation}
For simplicity we drop for the moment the site index $i$ and the
respective summation, that will be reintroduced at the end of the
following consideration. $L_1\dots L_4$ denote the different bands, and
$U_{L_1\cdots L_4}$ are the Coulomb matrix elements.
Restricting the electron scattering processes caused by the Coulomb
interaction to two involved subbands, only, we get instead of
(\ref{eq:H1}):
\begin{eqnarray}\label{eq:H1expanded}
  H_{1}=\frac{1}{2}\sum_{LL^{\prime}}\sum_{\sigma\sigma^{\prime}}
  &&\left\{U_{LL^{\prime}}c^{+}_{L\sigma}c^{+}_{L^{\prime}\sigma^{\prime}}
    c^{}_{L^{\prime}\sigma^{\prime}}c^{}_{L\sigma}\right.\nonumber\\
  &&\left.+J_{LL^{\prime}}c^{+}_{L\sigma}c^{+}_{L^{\prime}\sigma^{\prime}}
    c^{}_{L\sigma^{\prime}}c^{}_{L^{\prime}\sigma}\right.\nonumber\\
  &&\left.+J^{\star}_{LL^{\prime}}c^{+}_{L\sigma}c^{+}_{L\sigma^{\prime}}
    c^{}_{L^{\prime}\sigma^{\prime}}c^{}_{L^{\prime}\sigma}\right\}
\end{eqnarray}
In the case of Gd the band indices $L$ and $L^{\prime}$ can be
attributed either to a flat 4f band ($L\rightarrow f$) or to a broad
(5d/6s) conduction band ($L\rightarrow m$). In an obvious manner we can then split the
Coulomb interaction into three different parts,
\begin{equation}\label{eq:H1parts}
  H_1=H_{dd}+H_{ff}+H_{df},
\end{equation}
depending on wether both interacting particles stem from a conduction
band, $H_{dd}$, or both from a flat band $H_{ff}$, or one from a flat
band the other from a conduction band, $H_{df}$. The first term,
$H_{dd}$, refers to electron correlations in the broad conduction
bands. We consider them not to be decisive for the characteristic
Gd-physics. According to our concept (section \ref{sec:GProcedure}) $H_{dd}$ does
not enter explicitely our model being rather accounted for by the
single-particle energy renormalization (\ref{eq:HoppingRen}). $H_{ff}$ is
built by pure 4f correlations. The main influence of the 4f electrons on
the Gd-physics is due to the fact that they form permanent localized
magnetic moments. So $H_{ff}$ is unimportant as part of our
model-Hamiltonian and we are left with the interaction between localized
and itinerant electrons:
\begin{eqnarray}\label{eq:Hdfexpanded}
&H_{df}&=\sum_{mf\sigma\sigma^{\prime}}\bigl\{U_{mf}
c^{+}_{m\sigma}c^{+}_{f\sigma^{\prime}}c^{}_{f\sigma^{\prime}}
c^{}_{m\sigma}+
J_{mf}c^{+}_{m\sigma}c^{+}_{f\sigma^{\prime}}c^{}_{m\sigma^{\prime}}
c^{}_{f\sigma}\bigr.\nonumber\\
\bigl.&+&\frac{1}{2}J^{\star}_{mf}c^{+}_{m\sigma}c^{+}_{m\sigma^{\prime}}c^{}_{f\sigma^{\prime}}
c^{}_{f\sigma}+\frac{1}{2}J^{\star}_{fm}c^{+}_{f\sigma}c^{+}_{f\sigma^{\prime}}c^{}_{m\sigma^{\prime}}
c^{}_{m\sigma}\bigr\}
\end{eqnarray}
The last two terms do not contribute since the Gd$^{3+}$-4f shell has
its maximum spin $S=7/2$. All the seven 4f electrons have to occupy
different subbands and none of the seven subbands will be doubly
occupied. By use of the electron spin operator,
\begin{equation}
  \sigma^{+}=\hbar c^{+}_{\uparrow}c^{}_{\downarrow};\:
  \sigma^{-}=\hbar c^{+}_{\downarrow}c^{}_{\uparrow};\:
\sigma^{z}=\frac{\hbar}{2}(n_{\uparrow}-n_{\downarrow})\:
\end{equation}
($n_{\sigma}=c^{+}_{\sigma}c^{}_{\sigma}$) we get $H_{df}$ in the following compact form
\begin{equation}\label{eq:Hdffinal}
  H_{df}=-\frac{2}{\hbar^2}\sum_{mf}J_{mf}\boldsymbol{\sigma}_m
  \cdot\boldsymbol{\sigma}_f+\sum_{mf}\bigl(U_{mf}-\frac{1}{2}J_{mf}\bigr)n_{m}n_{f}
\end{equation}
with $n_{m(f)}=n_{m(f)\uparrow}+n_{m(f)\downarrow}$. For all processes of
interest the number of f electrons per site is fixed, $n_{f}$ is
therefore only a $c$-number. The last term in (\ref{eq:Hdffinal}) does not
really provide an fd-interaction. It leads only to a rigid shift of the
atomic levels being therefore fully accounted for by
the renormalization (\ref{eq:HoppingRen}) of the single-particle part of
the Hamiltonian. By defining the spin operator $\vec{S}$ of the local f
moment
\begin{equation}\label{eq:SpinOp}
  \vec{S}=\sum_{f}\boldsymbol{\sigma}_f,
\end{equation}
and by assuming that the interband exchange $J_{mf}$ is independent of
the special index-pair $m,f$
\begin{equation}\label{eq:Jmf}
  J_{mf}\equiv\frac{1}{2}J
\end{equation}
the interaction term reads after the reintroduction of the lattice site
dependence: 
\begin{eqnarray}\label{eq:Hsf}
  H_{df}&=&-\frac{J}{\hbar^2}\sum_{im}\boldsymbol{\sigma}_{im}\cdot\vec{S}_i\nonumber\\
  &=&-\frac{J}{2\hbar}\sum_{im\sigma}\left\{z_{\sigma}S^{z}_{i}n_{im\sigma}
    +S^{\sigma}_ic^{+}_{im-\sigma}c^{}_{im\sigma}\right\}
\end{eqnarray}
Here we have used the abbreviations
\begin{equation}\label{eq:abbrev}
  S^{\sigma}_{j}=S^{x}_{j}+iz_{\sigma}S^{y}_{j};\:
  z_{\sigma}=\delta_{\sigma\uparrow}-\delta_{\sigma\downarrow}
\end{equation}
The single-band version (non-degenerate s band) of (\ref{eq:Hsf}) is
well-known as the interaction part of the so-called Kondo-lattice model
(KLM)\cite{DON77}, in the older literature more appropriately denoted as
s-f or s-d model\cite{DDN98,NOL79,OVC91}. In the multiband case we have in
$H_{df}$ simply an additional summation over the orbital index $m$. The
first term of (\ref{eq:Hsf}) describes an Ising-like interaction of the
two spin operators, while the other provides spin exchange processes
between localized moment and itinerant electron. Spin exchange may
happen by three different elementary processes: Magnon emission by an
itinerant $\downarrow$-electron, magnon absorption by a
$\uparrow$-electron and also formation of a quasiparticle, which is
called \textit{``magnetic polaron''}. The latter can be understood as a
propagating electron \textit{``dressed''} by a virtual cloud of repeatedly
emitted and reabsorbed magnons corresponding to a polarization of the
immediate localized spin neighbourhood.

Our model-Hamiltonian, built up by the partial operators (\ref{eq:H0}) and
(\ref{eq:Hsf}),
\begin{equation}\label{eq:H}
  H=H_{0}+H_{df}
\end{equation}
can be considered as \textit{``multiband Kondo-lattice model''}
(m-KLM). While in the interaction part $H_{df}$ the multiband aspect
appears only as an additional summation, the subbands are intercorrelated
via the single-particle term $H_{0}$.

An important model parameter is of course the effective coupling
constant $JS/W$ where $W$ is the width of the \textit{``free''}
Bloch-band and $S$ the local spin value. It turns out that in particular
the sign of $J$ is decisive. Other model parameters are the lattice
structure and the band occupation
\begin{equation}\label{eq:total_n}
  n=\sum_{m\sigma}\langle n_{m\sigma}\rangle
\end{equation}
In case of an s-band $n$ is a number in between 0 and 2.
\subsection{Exact limiting case}\label{ssec:ExactLimitingCase}
\begin{figure}[ht]
\includegraphics[width=0.4\textwidth]{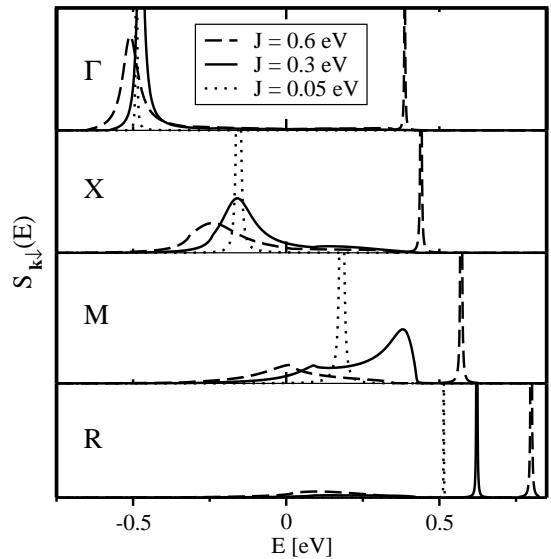}
\caption{Exact $\downarrow$ spectral density of the single-band Kondo-lattice model at
$T=0$ as function of the energy for several symmetry points in the first
Brillouin zone and for different exchange couplings $J$. Parameters:
$S=1/2$, $W=1$ eV, $n=0$, sc lattice.}
\label{fig:polaron}
\end{figure}

The many-body problem provoked by the model Hamiltonian (\ref{eq:H}) is
rather sophisticated, up to now not exactly solvable for the general
case.
Fortunately, however, there exists a non-trivial, very illustrative
limiting case which is rigorously tractable, nevertheless exhibiting all
the above mentioned elementary excitations
processes\cite{MSN01,SMA81,AED82}. It refers to a single electron in an
otherwise empty conduction band being coupled to a ferromagnetically
saturated moment system. Such a situation is met, e.g., for the
ferromagnetic semiconductor EuO at $T=0$. Because of the empty band and
the totally aligned spin system the hierarchy of equation of motions of
the single-electron Green function (\ref{eq:GF}, \ref{eq:GFelements})
decouples exactly. One can exploit exact relationships of the following
kind: 
\begin{eqnarray}\label{eq:ExactRel}
  \langle\dots c_{im\sigma}\rangle&=&\langle
  c^{+}_{im\sigma}\dots\rangle=0;\:
  \langle\dots S^{+}_{i}\rangle=
\langle S^{-}_{i}\dots\rangle=0\nonumber\\
\langle\dots S^z_{i}\rangle&=&\langle S^z_{i}\dots\rangle
=\hbar S\:\langle\dots\rangle
\end{eqnarray}
A troublesome but straightforward calculation then arrives at the
following result for the selfenergy matrix (\ref{eq:GF})
\begin{equation}\label{eq:MBMpolaron}
  {\hat M}^{}_{\vec{k}\sigma}(E)=-\frac{1}{2}z_{\sigma}JS\;{\hat{\rm I}}
+(1-z_{\sigma})\frac{\frac{1}{4}J^2S\;\frac{1}
{\hbar}{\hat G}_0(E+\frac{1}{2}JS)}{{\hat {\rm
  I}}-\frac{1}{2}J\frac{1}{\hbar}{\hat G}_0(E+\frac{1}{2}JS)}
\end{equation}
\begin{equation}
  \frac{1}{\hbar}{\hat G}_0(E)=\frac{1}{N}\sum_{\vec{k}}\left[(E+\mu)\hat{\rm
      I}-{\hat T}_{\vec{k}}\right]^{-1}
\end{equation}
The $\uparrow$ spectrum is especially simple because the $\uparrow$
electron cannot exchange its spin with the parallely aligned local spin
system. Only the Ising-type interaction in (\ref{eq:Hsf})
takes care for a rigid shift of the selfenergy by $-\frac{1}{2}J
S$. The spectral densities (\ref{eq:Skm}) are $\delta$-functions representing
quasiparticles with infinite lifetimes. Real correlation effects appear,
however, in the $\downarrow$ spectrum. The essentials can be seen
already for a non-degenerate s band. Fig.~\ref{fig:polaron} shows the
energy dependence of the $\downarrow$-spectral density
$S_{\vec{k}\downarrow}(E)$ for some symmetry points. Furthermore, we have chosen a sc lattice,
$S=1/2$ and $W=1$ eV. 

For weak coupling (e.g. $J=0.05$ eV) the spectral
density consists of a single pronounced peak. The finite width points to
a finite quasiparticle lifetime due to some spin flip processes, but the
sharpness of the peaks indicates a long living quasiparticle. This
changes drastically even for rather moderate effective exchange
couplings $JS/W$. One observes in certain parts of the Brillouin zone,
for strongly coupled systems even in the whole Brillouin zone, that the
spectral density splits into two parts. The sharp high-energy peak
belongs to the formation of the magnetic polaron while the broad
low-energy part consists of scattering states due to magnon emission by
the $\downarrow$ electron. As long as the polaron peak is above the
scattering spectrum the quasiparticle has even an infinite lifetime. The
scattering spectrum is in general rather broad because the emitted
magnon can carry away any wave-vector from the first Brillouin
zone. Because of the concomitant spinflip magnon emission can happen only if
there are $\uparrow$ states within reach. Therefore, the scattering part
extends just over that energy region where $\rho_{\uparrow}(E)\ne
0$. Sometimes, as e.g. for $J=0.6$ eV at the $\Gamma$ point
(Fig.~\ref{fig:polaron}), the scattering part is surprisingly bunched
together to a prominent peak, therefore certainly visible in
a respective (inverse) photoemission experiment. Note that the results
in Fig.~\ref{fig:polaron} are exact and free of any uncontrollable
approximation. They exhibit typical correlation effects which are by no
means reproduceable by a single-electron theory.

Very important for the following procedure is the simple $\uparrow$
result. It tells us that at ($T=0$, $n=0$) the df-exchange interaction
takes care only for a rigid shift of the total energy spectrum without
any deformation, which is therefore identical to the
\textrm{``free''} Bloch spectrum. Furthermore for this special case
a mean-field approximation turns out to be exact. Though not exactly provable,
many reliable approaches\cite{NRM97,MSN01,NRR01} show that this holds, at least to a good approximation, for
finite band occupation, too. This will be demonstrated in Fig.~\ref{fig:ASW-MCDA}
for the actual case of Gd.

\subsection{Band-structure calculations}\label{sub:BandCalc}
As described in section \ref{sec:GProcedure} the hopping integrals
$T^{mm^{\prime}}_{ij}$ in the single-particle Hamiltonian $H_{0}$
(\ref{eq:H0}) have to absorb the influences of all those
interactions which are not directly covered by our model Hamiltonian
(\ref{eq:H}). For this purpose we have performed a spin polarized
scalar-relativistic ASW-band calculation for ferromagnetic hcp Gd. The
result is plotted in Fig.~\ref{fig:GdBS}. The 4f electrons are
considered as valence electrons. In section \ref{sec:intro} we already
commented on the wrong
position of the 4f levels. However, we need
as input for the determination of the quasiparticle spectrum only the
$\uparrow$ part of the bandstructure because of the following
reason. The main problem, when using LDA-DFT results as renormalized
single-particle input, is to avoid a double counting of just the decisive
interband exchange coupling, namely once explicitely in the
interaction-part $H_{1}$ of the model-Hamiltonian and then once more
implicitely by the effective single-particle input. The exact limiting
case of the last section gives the hint how to circumvent this
double-counting problem. For an empty band ($n=0$) and ferromagnetic
saturation the $\uparrow$ spectrum is only rigidly shifted by a constant
energy amount ($-\frac{1}{2}JS$) compared to the \textit{``free''}
Bloch spectrum. As mentioned, model approaches convincingly demonstrate
that this remains true for finite band occupation (less than
half-filled bands !), too, at least to a very good approximation. For
less than half-filled bands we therefore can identify the ($T=0$,
$\sigma=\uparrow$)-LDA results (without the 4f part) with the free Bloch
energies. By this procedure we do not explicitely switch off the
df-interband exchange but rather exploit the fact that for the
mentioned special case the interband exchange leads only to a trivial
rigid shift. The wrong position of the LDA-4f $\downarrow$ states
therefore does not bother us because we need excludingly the $\uparrow$
spectrum. On the other hand, we can be sure that all the other
interactions, as e.g. the Coulomb interaction of the 5d electrons, are
excellently accounted for by the LDA. We note in passing, that for more
than half-filled bands we would have to
take the ($T=0$, $\sigma=\downarrow$)-spectrum (particle-hole symmetry).

After defining the single-particle input there remains only one
parameter, namely the exchange coupling $J$. It is not considered as a free
parameter, but taken from the bandstructure calculation. It is commonly
accepted that an LDA treatment of ferromagntism is quite compatible with
a mean-field ansatz\cite{JWI76,WKG79}, so that the exchange splitting in
Fig.~\ref{fig:GdBS} should amount to $\Delta=JS$ (see next section). We
realize, however, that the assumption of a rigid splitting is too simple.
A slight energy and wave-vector dependence of the exchange splitting is
found by LDA, too. We have therefore averaged the $T=0$-splitting over
$N_{p}$ prominent features in the Q-DOS of Gd arriving at 
\begin{equation}\label{eq:feature}
  J=\frac{1}{N_{p}S}\sum_{p}\Delta_{p}\approx 0.3\:\textrm{eV}
\end{equation}
This is of the same order of magnitude as found for Eu chalcogenides
\cite{SNO99a,MNO02}. There are then no other parameters in our theory.

The validity of the above assumptions will be demonstrated later with
Fig.~\ref{fig:ASW-MCDA}. 
\section{Many-body evaluation}\label{sec:MBeval}
The multiband-KLM (\ref{eq:H}) does not allow a rigorous solution except
for some special cases. Approximations are unavoidable. There are two
partial problems to be solved, one concernig the
ferromagnetism mainly provoked by the localized 4f moments, the other
dealing with the temperature reaction of the conduction-band states due to the
magnetic state of the moment system. In addition, both parts have to be
combined in a self-consistent manner.

For the electronic part we have developed in the past several approaches
\cite{NRM97,NRR01,NRR03}, which all lead, at least qualitatively, to the
same result. The \textit{``interpolating selfenergy
  approach''}\cite{NRR01,NRR03} is in particular trustworthy for almost
empty or almost full bands. For intermediate fillings as in the case of
Gd the \textit{``moment conserving decoupling approach''}
(MCDA)\cite{NRM97} seems to be more recommendable. So we use it
here. Since this approach has been exhibited in detail in
Ref.~\onlinecite{NRM97} we can restrict ourselves in the following to
the central parts which are vital for the understanding of the
underlying procedure. To make the representation as clear as possible we
develop the method in the next section for the special case of a
non-degenerate band. The orbital index is then superfluous. The
generalization for the degenerate case is straightforward. The
investigation of the magnetic part follows in the second subsection.
\subsection{Conduction electron selfenergy}
According to (\ref{eq:GF}) the solution of the problem needs the knowledge of
the selfenergy $M_{\vec{k}\sigma}(E)$. The above-mentioned MCDA is a
non-perturbational Green-function theory. It starts from the equation of
motion of the site-dependent single-electron Green function
(\ref{eq:GF}):
\begin{eqnarray}\label{eq:EOM}
\lefteqn{\sum_{m}\bigl((E+\mu)\delta_{im}-T_{im}\bigr)G_{mj\sigma}(E)=\hbar\delta_{ij}}\nonumber\\
&&\hspace{5em}-\frac{1}{2}J\Bigl(z_{\sigma}\Gamma_{ii,j\sigma}(E)+F_{ii,j\sigma}(E)\Bigr)
\end{eqnarray}
Our approximation attacks the equations of motion of the
\textit{``Ising-function''} 
\begin{equation}\label{eq:IsingGF}
  \Gamma_{im,j\sigma}(E)=\langle\langle S^z_{i}c^{}_{m\sigma};c^{+}_{j\sigma}\rangle\rangle_{E}
\end{equation}
and the \textit{``spin-flip function''}:
\begin{equation}\label{eq:spinFlipGF}
  F_{im,j\sigma}(E)=\langle\langle S^{-\sigma}_{i}c^{}_{m-\sigma};c^{+}_{j\sigma}\rangle\rangle_{E}
\end{equation}
These equations of motion contain still higher Green functions which
are decoupled to get a closed system of equations. Let us exemplify the
procedure by a \textit{``higher''} Green function of the type
$\langle\langle
A^{}_{i}[c^{}_{l\sigma},H_{df}]_-;c^{+}_{j\sigma}\rangle\rangle_{E}$, where
$A^{}_{i}$ is any combination of local-moment and band operators. The
off-diagonal terms $i\ne m$ are approached by use of the selfenergy
elements $M_{lr\sigma}$ (\textit{``selfenergy trick''}), in a certain
sense as a generalization of the exact equation (\ref{eq:Fsolution}):
\begin{equation}\label{eq:Setrick}
  \langle\langle
A^{}_{i}[c^{}_{l\sigma},H_{df}]_-;c^{+}_{j\sigma}\rangle\rangle_{E}
\Rightarrow \sum_{r}M_{lr\sigma}(E)\langle\langle A^{}_{i}c^{}_{r\sigma};c^{+}_{j\sigma}\rangle\rangle_{E}
\end{equation}
The right-hand side is a linear combination of \textit{``lower''} Green
functions with the selfenergy elements as selfconsistently to be
determined coefficients. To account for the strong local correlations
the diagonal terms $i=l$ are handled with special care:
\begin{eqnarray}\label{eq:GFDiagonal}
  \langle\langle
 A_{i}[c_{i\sigma},H_{df}]_-;c^{+}_{j\sigma}\rangle\rangle_{E}
&=&\alpha_{\sigma}G_{ij\sigma}(E)+\beta_{\sigma}\Gamma_{ii,j\sigma}(E)\nonumber\\
&+&\gamma_{\sigma}F_{ii,j\sigma}(E)
\end{eqnarray}
Such an ansatz is constructed in such a way that all known exact
limiting cases (atomic limit, ferromagnetic saturation, local spin
$S=1/2$, $n=0$, $n=2$,\dots) are exactly fulfilled. The at first unknown
coefficients $\alpha_{\sigma}$, $\beta_{\sigma}$, $\gamma_{\sigma}$ are
eventually found by equating exact high-energy expansion (spectral
moments) of the selfenergy. As the other above-mentioned
methods\cite{NRR01,NRR03} the MCDA
arrives at the following structure of the selfenergy:
\begin{equation}\label{eq:Selfenergy}
  M_{\vec{k}\sigma}(E)=-\frac{1}{2}Jz_{\sigma}\langle
  S^z\rangle+J^2D_{\vec{k}}(E; J) 
\end{equation}
Restriction to the first term, only, is just the mean-field approach to
the KLM, which is correct for sufficiently weak couplings $J$, being
mainly due to the Ising-part in (\ref{eq:Hsf}). Without the second part
it would give rise to a spin-polarized splitting of the conduction
band. The term $D_{\vec{k}\sigma}(E;J)$ is more complicated being
predominantly determined by spin exchange processes due to the spin-flip
term in the Hamiltonian (\ref{eq:Hsf}). It is a complicated functional
of the selfenergy itself, and that for both spin directions,
i.e. (\ref{eq:Selfenergy}) is an implicit equation for
$M_{\vec{k}\sigma}(E)$ and not at all an analytic
solution. $D_{\vec{k}\sigma}(E;J)$ depends, furthermore, on mixed spin
correlations such as $\langle S^z_in_{i\sigma}\rangle$, 
$\langle S^{+}_ic^{+}_{i\downarrow}c^{}_{i\uparrow}\rangle$,\dots,
built up by combinations of localized-spin and itinerant-electron
operators. Fortunately, all these mixed correlations can rigorously be
expressed via the spectral theorem by any of the Green functions
involved in the hierarchy of the MCDA. However, there are also pure
local-moment correlation functions of the form $\langle S^z_i\rangle$,
$\langle S^{\pm}_iS^{\mp}_i\rangle$, $\langle (S^{z}_i)^2\rangle$,\dots
which also have to be expressed by the electronic selfenergy
$M_{\vec{k}\sigma}(E)$.
\subsection{Modified RKKY interaction}\label{ssec:MRKKY} 
To get such expectation values of local-spin combinations we map the
interband exchange operator (\ref{eq:Hsf}) on an effective Heisenberg
Hamiltonian\cite{NRM97,SNO02}:
\begin{equation}\label{eq:Map}
  H_{df}=-\frac{J}{\hbar^2}\sum_{im}\boldsymbol{\sigma}_{im}\cdot\vec{S}_i 
  \Longrightarrow -\sum_{ij}J^{\textrm{eff}}_{ij}\;\vec{S}_i\cdot\vec{S}_j
\end{equation}
We use here again the full multiband version. The mapping is done by
averaging out the electronic degrees of freedom
$\boldsymbol{\sigma}_{im}\rightarrow
\langle\boldsymbol{\sigma}_{im}\rangle^{(c)}$. That means, in the last
analysis, to determine the expectation value 
$\langle
c^{+}_{\vec{k}+\vec{q}m\sigma}c^{}_{\vec{k}m\sigma^{\prime}}\rangle^{(c)}$
The averaging $\langle\cdots\rangle^{(c)}$ has to be done in the
conduction electron subspace where the local spins $\vec{S}_i$ can be
treated as classical variables:
\begin{equation}\label{eq:ResAverage}
  \langle
c^{+}_{\vec{k}+\vec{q}m\sigma}c^{}_{\vec{k}m\sigma^{\prime}}\rangle^{(c)}
=\frac{1}{\Xi^{\prime}}\textrm{Tr}(e^{-\beta
  H^{\prime}}c^{+}_{\vec{k}+\vec{q}m\sigma}c^{}_{\vec{k}m\sigma^{\prime}}) 
\end{equation}
$H^{\prime}$ is formally the same as in (\ref{eq:H}), except for the
fact that for the averaging process the f-spin operators are to be
considered as c numbers, therefore not affecting the
trace. $\Xi^{\prime}$ is the corresponding grand partition function. We
use the spectral theorem for the \textit{``restricted''} Green function,
\begin{equation}\label{eq:ResGF}
  G^{mm^{\prime}}_{\vec{k}\sigma^{\prime},\vec{k}+\vec{q}\sigma}(E)
=\langle\langle
c^{}_{\vec{k}m\sigma^{\prime}};c^{+}_{\vec{k}+\vec{q}m^{\prime}\sigma}\rangle\rangle_{E}
\end{equation}
to fix the expectation value (\ref{eq:ResAverage}). Eq.~(\ref{eq:ResGF})
stands for the usual definition (\ref{eq:GFelements}) of a retarded
Green function, only the averages have to be done in the Hilbert space of
$H^{\prime}$. The equation of motion of $\hat G$ reads (in matrix
representation with respect to the orbital indices $m$, $m^{\prime}$):
\begin{eqnarray}\label{eq:ResEOM}
&&  \hat
G_{\vec{k}\sigma^{\prime},\vec{k}+\vec{q}\sigma}(E)=\delta_{\vec{q},\vec{0}}\delta_{\sigma\sigma^{\prime}}
\hat G^{(0)}_{\vec{k}}(E)-\nonumber\\
&&-\frac{J}{2N}\sum_{i\sigma^{\prime\prime}\vec{k}^{\prime}}
\Bigl(e^{i(\vec{k}-\vec{k}^{\prime})\cdot\vec{R}_i}(\vec{S}_i\cdot
\boldsymbol{\sigma})_{\sigma^{\prime}\sigma^{\prime\prime}}\cdot\hat
G^{(0)}_{\vec{k}}(E)\hat
G^{}_{\vec{k}^{\prime}\sigma^{\prime},\vec{k}+\vec{q}\sigma}(E)\Bigr.\nonumber\\
&&\Bigl.+e^{i(\vec{k}^{\prime}-(\vec{k}+\vec{q}))\cdot\vec{R}_i}(\vec{S}_i\cdot\boldsymbol{\sigma})
_{\sigma^{\prime\prime}\sigma}\hat
G_{\vec{k}\sigma^{\prime},\vec{k}^{\prime}\sigma^{\prime\prime}}(E)
\hat G^{(0)}_{\vec{k}+\vec{q}}(E)\Bigr)  
\end{eqnarray}
This equation is exact and can be iterated up to any desired
accuracy. $\hat G^{(0)}_{\vec{k}}(E)$ is the Green function matrix of
the \textit{``free''} electron system:
\begin{equation}\label{eq:GF0}
  \hat G^{(0)}_{\vec{k}}(E)=\hbar\left[(E+\mu)\hat{\rm I}-\hat T_{\vec{k}}\right]^{-1}
\end{equation}
If we stop the iteration in (\ref{eq:ResEOM}) after the first nontrivial
step, i.e. replacing $\hat G$ on the right-hand side by the \textit{``free''} Green function
matrix, then we arrive at the well-known
RKKY-result\cite{SNO02}, which can be equivalently derived by use of
conventional second-order perturbation theory with respect to $J$
starting from the unpolarized conduction electron gas. To incorporate
the exchange-induced conduction electron spin polarization to a higher
degree we replace the restricted Green function on the right-hand side
of (\ref{eq:ResEOM}) not by the \textit{``free''} but by the full
single-electron Green function matrix $\hat G_{\vec{k}\sigma}(E)$
defined in (\ref{eq:GF}):
\begin{eqnarray}\label{eq:ResApprox}
  \hat G_{\vec{k}^{\prime}\sigma^{\prime\prime},\vec{k}+\vec{q}\sigma}(E)
  &\longrightarrow&
  \delta_{\vec{k}^{\prime},\vec{k}+\vec{q}}\delta_{\sigma^{\prime\prime}\sigma} 
  \hat G_{\vec{k}+\vec{q}\sigma}(E)\\
  \hat G_{\vec{k}\sigma^{\prime},\vec{k}^{\prime}\sigma^{\prime\prime}}(E)
  &\longrightarrow&
  \delta_{\vec{k},\vec{k}^{\prime}}\delta_{\sigma^{\prime}\sigma^{\prime\prime}} 
  \hat G_{\vec{k}\sigma^{\prime}}(E)
\end{eqnarray}
After some manipulations that replacement leads to the following
effective exchange integrals:
\begin{eqnarray}\label{eq:Effective}
  J^{\textrm{eff}}_{ij}&=&\frac{J^2}{8N\pi}\sum_{\vec{k}\vec{q}m\sigma}
e^{-i\vec{q}\cdot(\vec{R}_i-\vec{R}_j)}\int^{+\infty}_{-\infty}\:dE\:f_{-}(E)\star\nonumber\\
&&\Im\left[\left(\hat G_{\vec{k}\sigma}(E-\mu)\hat
    G^{(0)}_{\vec{k}+\vec{q}}(E-\mu)\right)^{mm}\right.\nonumber\\
&&+\left.\left(\hat G^{(0)}_{\vec{k}}(E-\mu)\hat
    G_{\vec{k}+\vec{q}\sigma}(E-\mu)\right)^{mm}\right]
\end{eqnarray}
These effective exchange integrals are functionals of the electronic
selfenergy $\hat M_{\vec{k}\sigma}(E)$ getting therewith a distinct
temperature - and carrier concentration dependence. Neglecting $\hat
M_{\vec{k}\sigma}(E)$, i.e. replacing in (\ref{eq:Effective}) the full
by the \textit{``free''} Green function, leads to the multiband version of the
conventional RKKY-exchange integrals\cite{KAS56,YOS57}. Via $\hat
M_{\vec{k}\sigma}(E)$ higher order terms of the conduction electron spin
polarization enter the \textit{``modified''} RKKY (\ref{eq:Effective})
which is therefore not restricted to weak couplings, only.

To get from the effective Heisenberg Hamiltonian (\ref{eq:Map}) the
magnetic properties of the multiband KLM we apply the standard
Tyablikow-approximation\cite{BTY59} which is known to yield convincing
results in the low as well as high temperature region. All the above
mentioned local-moment correlations are then expressed by the electronic
selfenergy. We therefore end up with a closed system of equations that
can be solved self-consistently for all quantities of
interest. For a detailed discussion of the so-found properties of the
single-band KLM the reader is referred to our previous
publications\cite{NRM97,SNO02,NMS03}. We use the theory in the next
section to find the electronic and the magnetic properties of the
ferromagnetic 4f-metal Gadolinium.
\section{MAGNETIC AND ELECTRONIC PROPERTIES OF GADOLINIUM}\label{sec:MEGd}
\begin{figure}[ht]
\includegraphics[width=0.45\textwidth]{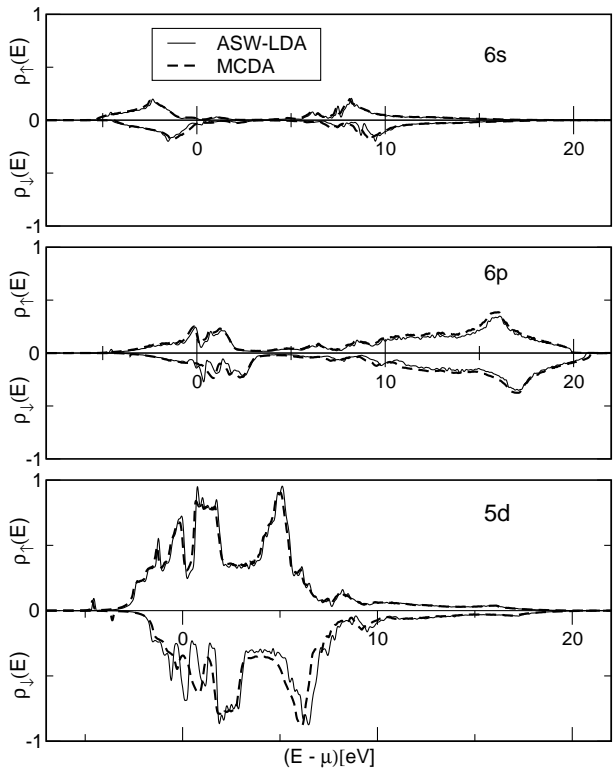}
\caption{Spin resolved densities of states of the 6s, 6p, and 5d bands of 
Gd as functions of the energy at $T=0$. Full lines for the ASW-LDA calculation, 
broken lines for our theory. $\uparrow$ ($\downarrow$) spectra in upper (lower) halves 
of the figures.}
\label{fig:ASW-MCDA}
\end{figure}

Fig.~\ref{fig:ASW-MCDA} shows the partial (5d, 6s, 6p) quasiparticle densities of states 
at $T=0$, as they are found by our method and compared to the pure ASW-LDA. The 
$\uparrow$ spin parts are almost identical for both methods. That confirms our 
procedure, explained in section \ref{sub:BandCalc}, for the 
combination of the many-body model evaluation and the "first principles" 
bandstructure calculation. Obviously a double counting of any decisive 
interaction has almost perfectly been avoided. The still observable very small 
deviations might be due to the finite band occupation. The statement 
that the up-spin spectrum 
at $T=0$ is only rigidly shifted (\ref{eq:MBMpolaron}) compared to the free spectrum can be 
proven, strictly speaking, only for empty bands. As mentioned, a lot of 
reliable approaches\cite{NRM97,SNO02,NRR03} support the assumption that this is 
true, at least to a very good approximation, for finite carrier 
densities, too. However, slight deviations may 
appear. Furthermore, the band- and wave-vector-independence of 
the exchange coupling $J$ (\ref{eq:feature}) is surely an oversimplification 
and may also contribute to the deviations in the $\uparrow$ spectrum. Nevertheless, the 
almost complete coincidence between LDA and model results demonstrate that 
there are hardly any exchange-caused 
correlation effects in the $\uparrow$ spectrum of the local-moment ferromagnet at 
$T=0$ (ferromagnetic saturation).

The $\downarrow$ part of the $T=0$-spectrum, however, exhibits already strong 
correlation effects due to the exchange coupling of the band states to 
the 4f moment system, predominantly in the 5d subband. They follow from 
magnon emission processes of the down-spin electrons and to a lesser 
extent from the formation of magnetic polarons.

Integration up to the Fermi edge yields the $T=0$ contribution 
of the conduction electrons to the magnetic moment. We find
\begin{equation}\label{eq:DmuB}
  \Delta\mu = 0.71\:\mu_{\rm B}.
\end{equation}
Since in our model the 4f moments have a fixed value of 7 $\mu_{\rm B}$ the total 
moment amounts to 7.71 $\mu_{\rm B}$ very close to the experimental value 
of 7.63 $\mu_{\rm B}$\cite{NLS63}. Our value is a bit smaller than that from the LDA+$U$ 
calculation in Ref.~\onlinecite{Kurz02}.

The procedure explained in the preceding sections allows for 
a determination of the full temperature dependence of the energy 
spectrum and the magnetic properties of Gd. The selfconsistent 
evaluation yields a ferromagnetic low-temperature phase with astonishing 
\begin{figure}[ht]
\includegraphics[width=0.45\textwidth]{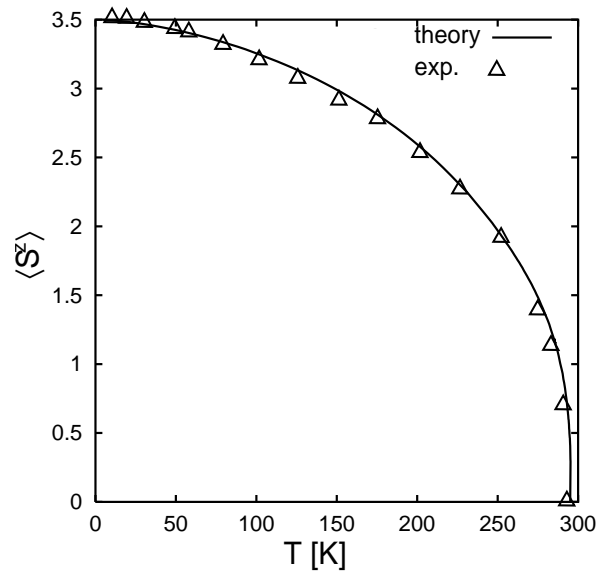}
\caption{Magnetization curve for ferromagnetic Gadolinium as function of
the temperature. The triangles represent experimental data taken from 
Ref.~\onlinecite{NLS63}.}
\label{fig:SzGd}
\end{figure}
precise key-data. Fig.~\ref{fig:SzGd} exhibits the magnetization curve in comparison 
to the experimental data of Ref.~\onlinecite{NLS63}. There is practically an exact 
agreement. In particular the calculated Curie temperature,
\begin{equation}\label{eq:TcCalc}
  T_{\rm C} = 294.1\:{\rm K},
\end{equation}
being known as a very sensitive entity of magnetism, hardly deviates 
from the experimental value of 293.2 K. Note that there is in principle 
no fitting parameter in our theory, even the exchange constant $J$ (\ref{eq:feature}) 
is taken from the LDA input. We therefore have to conclude that the 
modified RKKY theory (Section \ref{ssec:MRKKY}), with the effective exchange 
integrals being functionals of the conduction electron selfenergy, describes 
the ferromagnetism of Gd in an absolutely convincing manner.

\begin{figure}[h]
\includegraphics[width=0.45\textwidth]{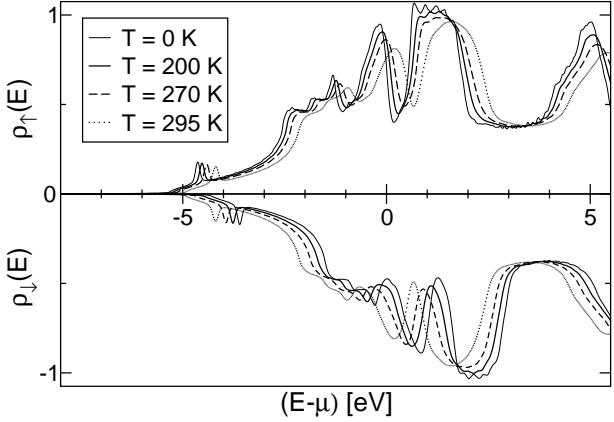}
\caption{Quasiparticle density of states of the valence and conduction 
bands of Gd as function of the energy $(E-\mu)$ ($\mu$: chemical potential) for
four different temperatures. The total densities of states consist 
of 5d, 6s, and 6p contributions.}
\label{fig:QDOST}
\end{figure}

Since we did not consider a direct exchange interaction between 
the localized 4f moments the induced spin polarization of the conduction 
electrons mediates the indirect coupling. The "a priori" only slightly 
correlated 5d/6s/6p band states therefore exhibit a distinct temperature 
dependence as can be seen for the total quasiparticle density of states 
in Fig.~\ref{fig:QDOST}. The $T=0$ splitting is responsible for the 
band contribution (43) 
to the total magnetic moment. With increasing temperature the induced 
splitting reduces steadily collapsing at $T_{\rm C}$. 
The shift is not at all rigid ("Stoner like"), but with clear 
deformations. The latter point to a substantial influence of nonlinear 
effects such as magnon emission and absorption and magnetic polaron 
formation, in particular what concerns the d states. The lower edge 
of the $\uparrow$ spectrum, predominantly built up by 6s states (Fig.~\ref{fig:ASW-MCDA}), shows 
a red shift upon cooling below $T_{\rm C}$ as it is typical for local-moment 
systems, first observed for insulators and semiconductors such as 
EuO and EuS\cite{BAT75}. The temperature behaviour at the 
chemical potential is not so clear.

\begin{figure}[h]
\includegraphics[angle=-90, width=\linewidth]{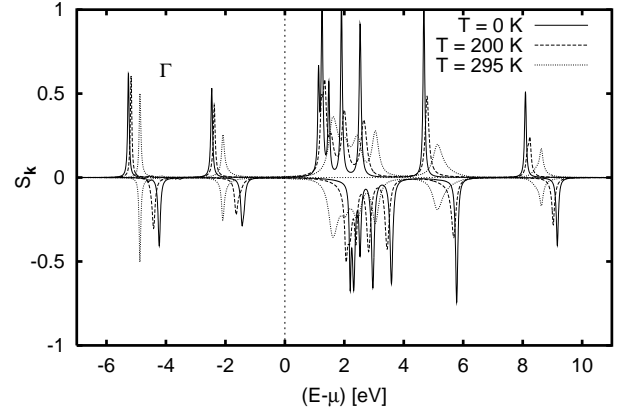}
\caption{Spin resolved single-electron spectral density of Gd at the
$\Gamma$ point as function of the energy ($E-\mu$) for three different temperatures. Upper 
half: $\uparrow$ spectrum, lower half: $\downarrow$ spectrum.}
\label{fig:SkG}
\end{figure}

\begin{figure}[ht]
\includegraphics[angle=-90, width=\linewidth]{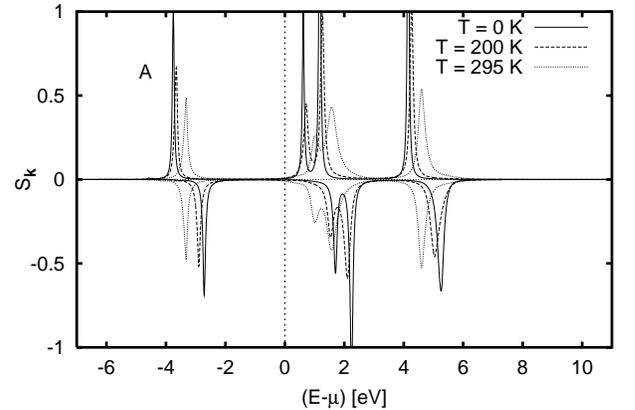}
\caption{The same as in Fig.~\ref{fig:SkG} but for the A point.}
\label{fig:SkA}
\end{figure}

\begin{figure}[ht]
\includegraphics[angle=-90, width=\linewidth]{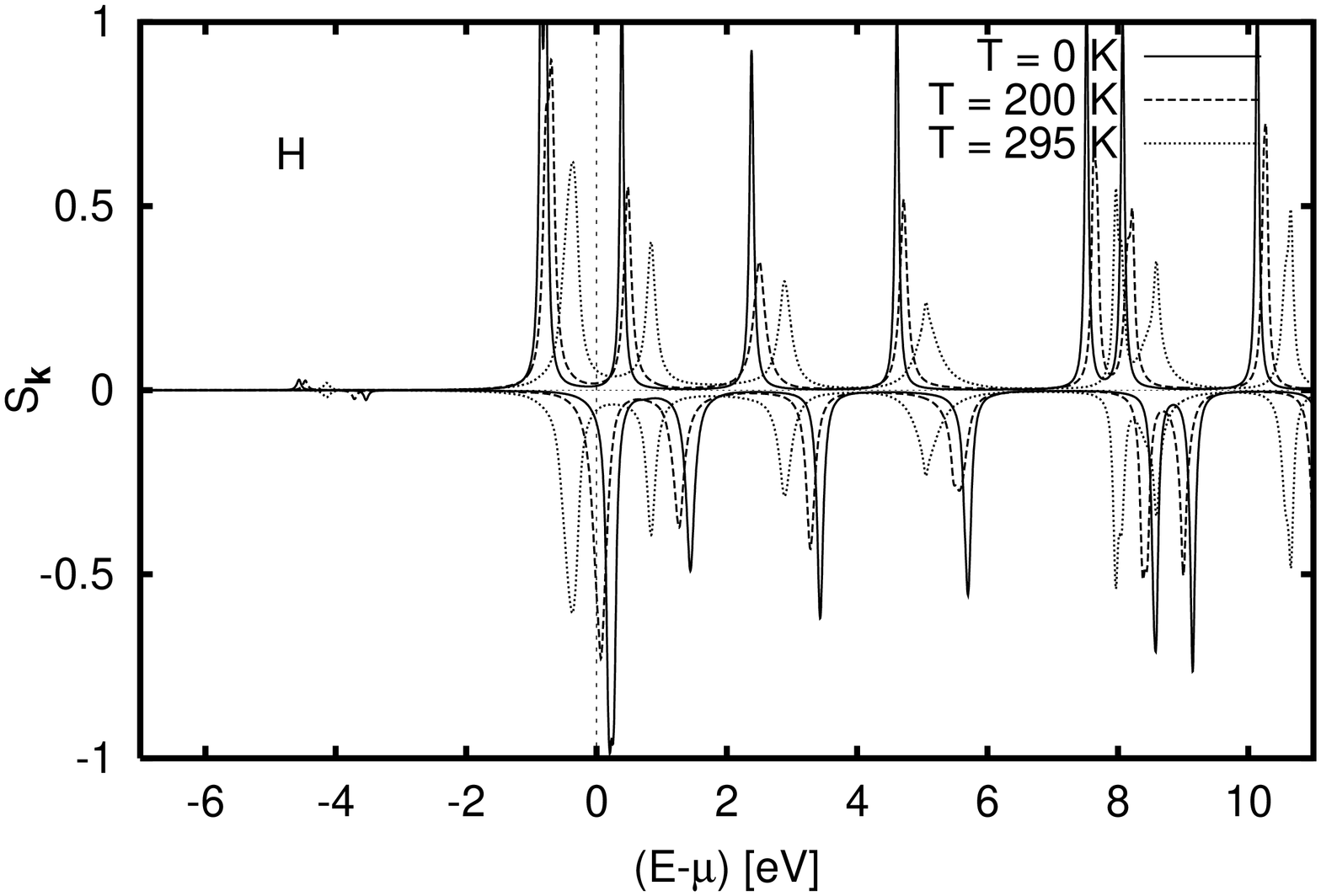}
\caption{The same as in Fig.~\ref{fig:SkG} but for the H point.}
\label{fig:SkH}
\end{figure}

\begin{figure}[ht]
\includegraphics[angle=-90, width=\linewidth]{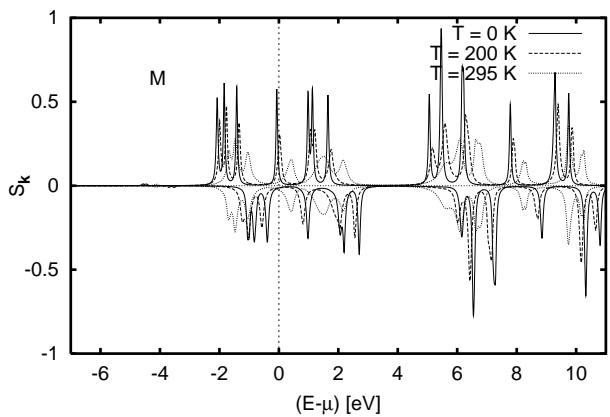}
\caption{The same as in Fig.~\ref{fig:SkG} but for the M point.}
\label{fig:SkM}
\end{figure}

The single-electron spectral density (\ref{eq:Skm}) represents the bare line shape
of an angle- and spin-resolved photoemission experiment. Pronounced
peaks in the spectral density define the quasiparticle band structure. For
four high-symmetry points ($\Gamma$, A, H, M) we have calculated the
energy dependence of the spectral density in the valence and conduction band region. 
The results for three different temperatures ($T=0$, 200, 295 K) are 
represented in Figs.~\ref{fig:SkG} to \ref{fig:SkM}. The $T=0$-$\uparrow$ 
spectra always consist of relatively 
sharp peaks pointing at quasiparticles with long, sometimes even infinite
lifetimes. In case of infinite lifetime (real selfenergy) the spectral 
density is a $\delta$-function. For plotting reasons we have then added a small
imaginary part ($i\Delta$; $\Delta=0.01$) to the electronic selfenergy. For empty energy
bands the $\uparrow$ spectrum would consist at $T=0$ exclusively of $\delta$-peaks. This 
is just the exact limiting case discussed in
Sec.~\ref{ssec:ExactLimitingCase}. It 
means nothing else than that a $\uparrow$ electron cannot undergo 
any scattering process if the
localized 4f moments are ferromagnetically saturated. However, for finite
and not fully spinpolarized band occupations some spin exchange processes 
may happen giving rise to slight quasiparticle dampings.
Fig.~\ref{fig:SkG} shows the energy dependence of the spectral density
at the $\Gamma$ point
for three different temperatures. $T=0$ means ferromagnetic 4f
saturation (Fig.~\ref{fig:SzGd}) while $T=295$ K is slightly above the calculated Curie
temperature (\ref{eq:TcCalc}). At $T=200$ K the moment system is partially ordered. The low energy
peaks belong to 6s states (Figs.~\ref{fig:GdBS} and \ref{fig:ASW-MCDA}). They are spin split in the 
ferromagnetic phase, where the induced exchange splitting diminishes 
continuously with increasing temperature, collapsing at $T=T_{\rm C}$ ("Stoner-like" 
behaviour). That agrees with the photoemission data of Kim et al.\cite{KAE92}. Similar 
temperature behaviour is found for the other quasiparticle peaks, 
too, and also for the other symmetry points A, H and M 
(Figs.~\ref{fig:SkA}, \ref{fig:SkH}, \ref{fig:SkM}).
These theoretical results contradict a bit our previous investigation\cite{REN99}
according to which in some cases a persisting splitting in the
paramagnetic phase should be possible. The ambiguity comes along with
the necessary 
decomposition of the total spectrum into non-degenerate subbands. That
can be done, in principle, in different ways, and, at least in our 
opinion, it is not ``a priori'' clear which is the correct procedure. In
this work we have used a method that retains the full atomic-orbital symmetry. 
The resulting rather broad subbands (Fig.~\ref{fig:ASW-MCDA}) cause correspondingly small 
effective exchange couplings $J/W$. The selfenergy
$M_{\vec{k}\sigma}(E)$ (\ref{eq:Selfenergy}) is then
dominated by the first term and therewith relatively close to the
mean-field solution of the sf- (Kondo-lattice-) model. The "Stoner
collapsing" is a typical feature of the weak-coupling (mean-field) region. The
band decomposition used in Ref.~\onlinecite{REN99} leads to substantially smaller subband
widths and therewith to stronger effective exchange couplings.

A general observation is that with increasing temperature the excitation
peaks are getting broader, i.e. quasiparticle lifetimes
decrease. Raising
temperature means enhancing magnon densities and therewith a higher 
probability for electron-magnon spinflip scattering. The d-like states near
and around the chemical potential $\mu$ exhibit stronger correlation effects 
than the low-energy 6s states or high-energy 6p states, again due to the 
larger effective coupling constant $J/W$. Contrary to the H and M points
(Figs.~\ref{fig:SkH}, \ref{fig:SkM}) there is no spectral weight at the chemical potential for
the $\Gamma$ and A points (Figs.~\ref{fig:SkG}, \ref{fig:SkA}).
\section{summary}
In this paper we have used a combination of a many-body approach to the 
Kondo-lattice (s-f) model with an LDA-DFT band structure calculation to
get in a realistic and selfconsistent manner the electronic and magnetic 
properties of the rare earth metal gadolinium. The many-body approach has 
previously been developed and tested in several model studies. It consists
of a moment-conserving decoupling approach for the single-electron Green
function, which fulfills a maximum number of exact limiting cases, and a
modified RKKY theory for the localized moment system. The effective 
exchange integrals between the localized spins turn out to be functionals of 
the electronic selfenergy. In the weak coupling limit the approach agrees
with the conventional RKKY theory.

As single-electron (Bloch) energies we have used the results of
an ASW band structure calculation therewith guaranteeing that all those
interactions 
which are not explicitly covered by the Kondo-lattice model are taken
into account in a rather realistic manner. An exact limiting case of the
model could be exploited to avoid the well-known double counting
problem. In a strict sense the method does not contain any really free
parameter. 
The 4f-5d exchange coupling constant $J$, which enters the theory via 
the Kondo-lattice model, is fitted by the LDA input.

The results of our theoretical investigation agree astonishingly
well with the experimental data of Gd. The selfconsistent approach predicts 
correctly a ferromagnetic low-temperature phase. The magnetic $T=0$
moment is
with 7.71 $\mu_{\rm B}$ very close to the experimental value of 7.63
$\mu_{\rm B}$. Even the 
extremely sensitive Curie temperature hardly deviates from the real Gd value
(theory: 294.1 K, experiment: 293.2 K). The valence and conduction bands 
exhibit a remarkable induced temperature dependence. The $T=0$ exchange 
splitting explains the excess moment of 0.63 $\mu_{\rm B}$ (or 0.71
$\mu_{\rm B}$), that cannot be 
ascribed to the seven 4f electrons. The temperature dependence of the 
exchange splitting roughly scales with the macroscopic magnetization
collapsing 
at $T_{\rm C}$ (``Stoner-behaviour'') as it has been observed in photoemission 
experiments. Correlation effects lead to a distinct temperature dependence
of the quasiparticle damping.

We believe that the proposed combination of a careful many-body
treatment of a proper theoretical model with an ab initio band structure
calculation yields a rather realistic description of the ferromagnetic 4f metal Gadolinium.
\section*{ACKNOWLEDGMENT}
Financial support by the ``Deutsche Forschungsgemeinschaft'' within the
SFB 290 (``Metallische d{\"u}nne Filme: Struktur, Magnetismus und
elektronischen Eigenschaften'') and the SFB 484 (``Kooperative
Ph{\"a}nomene im Festk{\"o}rper: Metall-Isolator-{\"U}berg{\"a}nge 
und Ordnung mikroskopischer Freiheitsgrade``) are gratefully acknowledged.
 
\end{document}